\documentclass[a4paper]{jpconf}
\usepackage{graphicx}
\begin{document}

\title{Neutrino oscillations beyond the Standard Model}

\author{F. del Aguila$^1$, J. Syska$^2$, M. Zra{\l}ek$^2$}

\address{$^1$ Departamento de F{\'\i}sica Te\'orica
y del Cosmos and CAFPE, Universidad de Granada, E-18071 Granada, Spain}
\address{$^2$ Institute of Physics, 
University of Silesia, Uniwersytecka 4, 40-007 Katowice, Poland}

\ead{marek.zralek@us.edu.pl}

\begin{abstract}
We address the possible impact of New Physics on neutrino 
oscillation experiments. 
This can modify the neutrino production, propagation and/or 
detection, making the full cross section non-factorizable in 
general.  
Thus, for example, the neutrino flux may not be properly 
described assuming an unitary MNS matrix and/or neutrinos 
may propagate differently depending of their Dirac or Majorana 
character. 
Interestingly enough, present limits on New Physics still 
allow for observable effects at future neutrino experiments.   
\end{abstract}
\section{Introduction}
In order to describe the possible impact of New
Physics (NP) in the full neutrino oscillation process 
we must properly convolute the production, 
propagation and detection amplitudes because 
in contrast with the minimal Standard Model (SM), 
these subprocesses do not in general factorize for 
relativistic neutrinos. 
Comparing with the general expression for 
neutrino oscillations we can learn: (i) Under which 
conditions the production and detection processes 
can be described by pure states as defined by the 
Maki-Nakagawa-Sakata mixing matrix, or we must use 
the density matrix formalism. (ii) How good the 
factorization approximation is. (iii) If we can 
only consider NP effects in the neutrino propagation, 
neglecting departures from the SM in the production 
and detection processes, or viceversa. 
We find using a general effective Lagrangian 
including charged and neutral scalar and vector 
interactions for both neutrino chiralities, and 
constraining the new couplings to satisfy present 
bounds on muon decay and neutron beta decay, that future  
more precise oscillation experiments can be sensitive 
to NP.  
 
\section{Main differences between the standard and NP approach}
Neutrinos $\nu_{i}$ with mass $m_{i}$ produced in the process 
\ $l_{\alpha}(\lambda_{\alpha}) + A(\lambda_{A})
\rightarrow \nu_{i}(\lambda) + B(\lambda_{B})$ 
can be described by the density matrix 
(see Ref. \cite{OSZ} for notation):
\begin{eqnarray}
\label{partial density in prod} 
\varrho^{\alpha}_{P} (\vec{p};
\lambda, i; \lambda', i') = \frac{1}{N_{\alpha}} \sum_{\lambda_{B}
\lambda_{A} \lambda_{\alpha}} A_{i}^{\alpha} (\vec{p}, \lambda;
\lambda_{B}, \lambda_{A}, \lambda_{\alpha}) \, A_{i'}^{\alpha *}
(\vec{p}, \lambda'; \lambda_{B}, \lambda_{A}, \lambda_{\alpha}).
\end{eqnarray}
This is equivalent to the usual description with pure 
Maki-Nakagawa-Sakata (MNS) states 
$|\nu_{\alpha}\rangle  = \sum_{i = 1}^{3} U^{*}_{\alpha
i}|\nu_{i}\rangle$ 
if neutrinos have SM interactions and are relativistic, 
depending the difference between both descriptions on the 
NP strength. 
Present bounds on NP allow for effects at the few per cent 
level, thus at the reach of future, more precise neutrino 
oscillation experiments \cite{OSZ}. 

Then, the production density matrix evolves along the 
distance $L = T$ to the detector with the Hamiltonian $\cal{H}$ 
describing the neutrino interaction with the surrounding medium, 
\begin{eqnarray}
\label{evolution of rho } \varrho_{P}^{\alpha}(\vec{p},  L=T=0,)
\rightarrow \rho_{D}^{\alpha}(\vec{p} ,L=T\neq0) = 
e^{-i {\cal{H}} \,T } \; \varrho^{\alpha}(\vec{p}, L=T=0) \; 
e^{i {\cal{H}} \,T } .
\end{eqnarray}
This generalizes the usual SM description \cite{Raffelt,Bergmann}. 
For instance, the different propagation through matter of Dirac and 
Majorana neutrinos in the presence of new right-handed interactions, 
which could be eventually observable, can be properly taken into account 
\cite{Paco}.

Finally, we can calculate the full cross section for the 
detection of the neutrino flavour $\beta$ in the process 
$\nu_{k} + C \rightarrow l_{\beta} + D$ from the density 
matrix at the detector $\varrho_{D}^{\alpha}$:  
\begin{eqnarray}
\label{cross sec} \sigma_{\alpha\rightarrow\beta}(E,L) &=&
\frac{p_{f}}{32 \ \pi \ s \ p_{i}}\frac{1}{(2 \  s_{C} + 1)}
\sum_{spins} \int d Lips \ A^{\beta}
\varrho^{\alpha}_{D}({\vec{p}},L=T) \ A^{\beta *},
\end{eqnarray}
where $E$ is the energy of the initial neutrino and $A^{\beta}$ 
the detection process amplitude \cite{Syska}.  
This cross section factorizes for relativistic neutrinos 
within the SM, $\sigma_{\alpha\rightarrow\beta}(E,L) = 
P_{\alpha\rightarrow\beta}(E,L) \  \sigma_{\beta}(E)$, 
where $P_{\alpha\rightarrow\beta}(E,L)$ is the corresponding 
oscillation probability and $\sigma_{\beta}(E)$ the 
detection cross section, but not in general.

Hence, generically the number of neutrinos of flavour $\beta$ 
produced per unit flux of neutrinos of flavour $\alpha$ and 
per scattering center is given by Eq. (\ref{cross sec}). 
Obviously, the size of the new effects depends on the strength 
of the NP. 
If we try to be as much model independent as possible and only 
consider the experimental constraints derived from muon decay 
and neutron beta decay (see {\it e.g.} \cite{Bounds}), we 
find that future more precise neutrino oscillation experiments 
can be sensitive to NP.  
As an example we plot in the Figure ({\it left}) the possible 
departure from the SM prediction for the 
$\nu_{\mu}\rightarrow\nu_{e}$ transition through the Earth 
for Dirac and Majorana neutrinos, and for $L =$ 732 km as a 
function of the incoming neutrino energy $E$ \cite{ASZZ}.   
Similarly for $\nu_{\mu}\rightarrow\nu_{\tau}$ in the 
Figure ({\it right}) but for propagation in vacuum, 
and for $E =$ 10.5 GeV as a function of the baseline distance $L$.  
\begin{figure}[!h]
\vspace{-0.2cm}
\hspace{-0.8cm}
\includegraphics[width=3.3in,height=5.9cm]{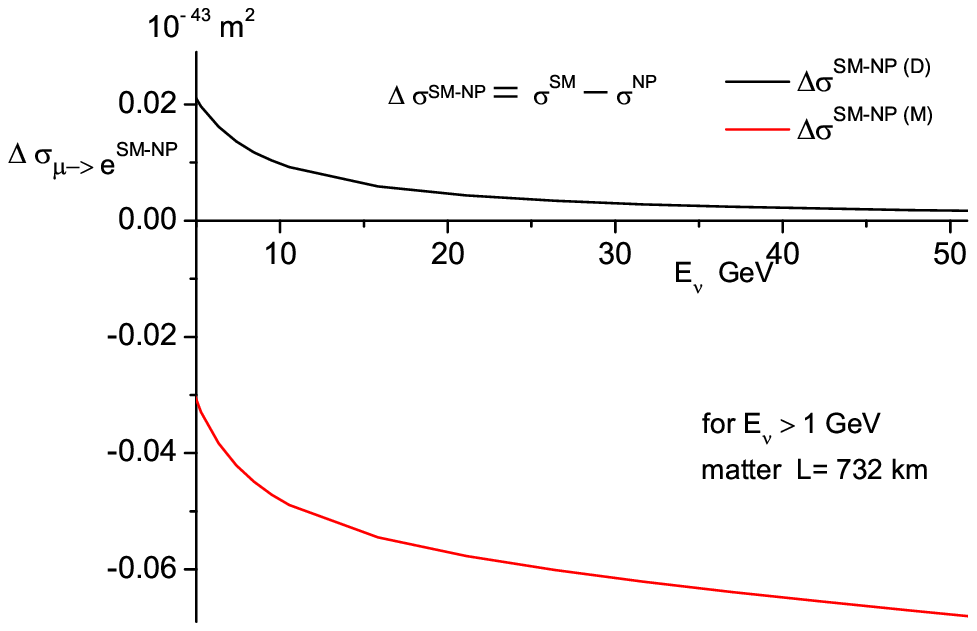} \hfill{}
\includegraphics[width=3.5in,height=6.2cm]{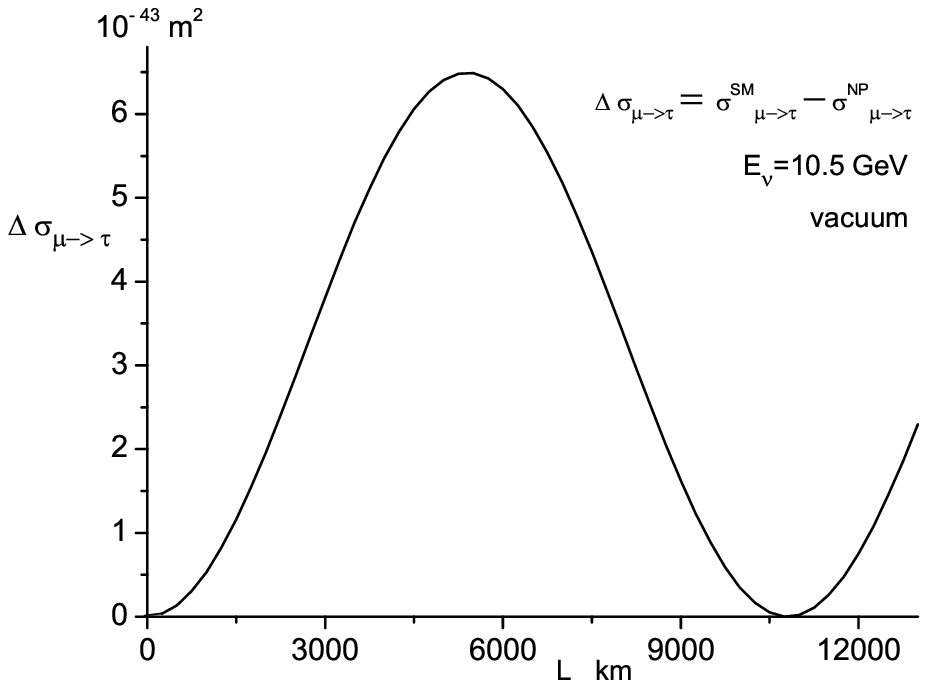} 
\label{figure}
\end{figure}

\end{document}